\documentclass{article}

\usepackage{PRIMEarxiv}

\usepackage[utf8]{inputenc} % allow utf-8 input
\usepackage[T1]{fontenc}    % use 8-bit T1 fonts
\usepackage{hyperref}       % hyperlinks
\usepackage{url}            % simple URL typesetting
\usepackage{booktabs}       % professional-quality tables
\usepackage{amsfonts}       % blackboard math symbols
\usepackage{nicefrac}       % compact symbols for 1/2, etc.
\usepackage{microtype}      % microtypography
\usepackage{lipsum}
\usepackage{fancyhdr}       % header
\usepackage{graphicx}       % graphics
\graphicspath{{media/}}     % organize your images and other figures under media/ folder

\title{SODA: a TypeScript/JavaScript Library for Visualizing Biological Sequence Annotation}

\author{ 
    {\hspace{1mm}Jack W.~Roddy}\\
	Department of Computer Science\\
	University of Montana\\
	Missoula, MT\\
	\texttt{jack.roddy@umontana.edu} \\
	\And
	{\hspace{1mm}George T.~Lesica} \\ 
	Department of Computer Science\\
	University of Montana\\
	Missoula, MT\\
	\texttt{george.lesica@umontana.edu} \\
	\And
	{\hspace{1mm}Travis J.~Wheeler} \\
	Department of Computer Science\\
	University of Montana\\
	Missoula, MT\\
	\texttt{travis.wheeler@umontana.edu} \\
}

\begin{document}
\maketitle

\begin{abstract}
We present SODA, a lightweight and open-source visualization library for biological sequence annotations that enables straightforward development of flexible, dynamic, and interactive web graphics.
SODA is implemented in TypeScript and can be used as a library within TypeScript and JavaScript.
\end{abstract}

\section{Introduction}
Annotation of biological sequences and the visualization of these annotations is central to molecular biology.
It is common to represent features (e.g. genes) annotated in the context of a reference sequence (e.g. a genome) using a combination of glyphs (rectangles, lines, arrows, etc.) and plots (line plots, bar plots, heatmaps, etc.).
Countless software tools have been developed that produce such visualizations, with modern applications typically developed using web technologies.
Most prominent among these are genome browsers such as JBrowse~\cite{buels2016jbrowse}, and hosted services such as Ensembl~\cite{howe2021ensembl} and the UCSC genome browser~\cite{navarro2021ucsc}.
Recent innovations include the grammar-based JavaScript toolkit for visualizing genomics data, Gosling~\cite{lyi2021gosling}.
Libraries also exist for visualizing annotation specific to protein sequences, such as ProtVista~\cite{watkins2017protvista} and its successor Nightingale.

These frameworks provide extensive visualization functionality, but may not meet the needs that arise in the development of custom visualization systems. 
Custom visualizations may require visual features or data types
not supported by out-of-the-box solutions, may involve interactive response to user actions that exceed the options provided by these frameworks, and may demand integration of multiple visual facets into a coherent whole. Furthermore, the use of standardized browsers may impose hefty dependencies and inclusion of unwanted GUI components.
To avoid these compromises,  developers of custom visualization systems often turn to core web browser technologies (HTML, CSS, SVG, Canvas, WebGL), or libraries that lightly abstract over those core technologies (D3\cite{bostock2011d3}, PixiJS\cite{pixijs}).

These developers would be well served by a front-end software library that enables extensive control over the form and function of annotation visualization elements useful for representing genome annotations, while further abstracting over most of the complexity of low-level libraries.
To our knowledge, there exists no lightweight web-based library that meets this need.
Here, we present SODA (Soda Obediently Draws Annotations), an open-source TypeScript/JavaScript library that aims to facilitate the development of flexible, dynamic, and interactive annotation visualization systems.
SODA is designed to enable augmentation with lower-level visualization libraries where needed, and it can also be used to generate simple one-time-use figures.
The SODA source code is available at \url{https://github.com/sodaviz/soda}, the library is packaged through NPM (@sodaviz/soda), and the website \url{https://sodaviz.org/} provides documentation and examples of visuals implemented with SODA.

% **************************************************************
% Keep this command to avoid text of first page running into the
% first page footnotes
\enlargethispage{-65.1pt}
% **************************************************************

\section{MATERIALS AND METHODS}

\subsection{Design}
SODA is a lightweight, purely front-end library implemented in TypeScript.
TypeScript is compiled to JavaScript, which means SODA can be used from either TypeScript or JavaScript.
Under the hood, SODA figures are created with Scalable Vector Graphics (SVG).
SODA is an object oriented library with few third-party dependencies.
SODA objects are designed with extension in mind, and they expose functions that allow for a great deal of programmatically-driven change at runtime.
Currently, SODA is focused on the presentation of annotations in linear context, with light support for representation in a circular context (e.g. for bacterial genomes).

SODA is a toolkit with which a developer may \emph{create} a visualization system, rather than a visualization system in and of itself.
The main consequence of this philosophy is that SODA does not produce visualizations using templating mechanisms commonly found in genome browser tools and the like.
Instead, SODA places fine-grained control in the hands of developers with a modest complexity trade-off: developers must define simple data structures to house their data, and functions that utilize SODA’s rendering API to produce a visualization. 

\subsection{High-level description of developing with SODA}
Development using SODA is largely focused on the configuration and management of Chart objects, which are wrapper objects that control SVG-based viewports in web pages. Creating a SODA visualization typically involves implementing simple objects that describe the annotation data being visualized, designing a data rendering payload that contains collections of those objects, and using the SODA API to configure the visual representation of the payload and the ways in which that visualization will respond to user inputs (e.g. clicking, panning, zooming).

\subsection{SODA’s treatment of data}
SODA is designed around a minimal abstraction of annotation data, rather than specific annotation data formats (e.g. BED, GFF3, GTF).
For gene-like annotations that describe an interval, this is a simple object composed of an identifier string, a start coordinate, and an end coordinate.
For annotations that describe position specific information throughout an interval, SODA uses the same object specification with an additional array/vector of position specific values.
All of the core SODA rendering features are designed to produce a visual representation of any JavaScript objects that extend this simple pattern.
In practice, annotations are commonly augmented with auxiliary data (e.g. alignment strand or score) which, during visualization, can be used to modulate some aspect of the visual representation of the annotation (e.g. color or opacity).
With this in mind, SODA rendering features are also designed such that it is easy for a developer to use additional object data fields to control glyph styling parameters.
Additionally, SODA provides suitable object definitions for several common annotation data formats along with accompanying parsing routines.

\subsection{SODA’s rendering API}
To render glyphs with SODA, Annotation objects are passed into rendering functions along with a configuration that specifies the target Chart object and styling parameters. For each style parameter, either a static value or a callback function may be provided. The callback functions are evaluated for each individual glyph, and the represented Annotation object and target Chart are passed as arguments.
Styling callback functions provide a simple mechanism to control glyph style using annotation data and may be re-evaluated at runtime to achieve visualization dynamics (e.g. zooming and transforming).

\section{Results}

Development of SODA was motivated by a need in our group to create genome-oriented visualizations for which existing genome visualization tools either were unable to provide desired functionality or were deemed unnecessarily heavyweight.
Here, we present several of these visualizations, with the goal of demonstrating SODA's flexibility in providing options for rendering and interactivity.
Working demonstrations for each example can be found at \url{https://sodaviz.org}, and the underlying source code can be found at \url{https://github.com/sodaviz/}.
These examples present applications of SODA for visualizing genome annotations, but it can be used to visualize protein annotations as well.

\subsection{Dfam visualization}
Dfam~\cite{storer2021dfam} is an open access database of Transposable Elements (TEs).
One feature on the Dfam website allows users to view a representation of the set of annotated TE instances in a relatively short region (up to 100,000 nucleotides) of a chromosome. In collaboration with the maintainers of Dfam, we replaced the original implementation with a more feature-rich variant using SODA (Fig.~\ref{fig:rmsk-dfam}).

The bulk of the annotations underlying the Dfam visualization are the result of comparing a genome to a database of known TE families. The resulting TE annotations are supplemented with annotations for instances of simple tandem repeats (repetitive sequence such as `atgatgatgatg'). In the linear genome portion of the visualization (top of Fig.~\ref{fig:rmsk-dfam}), each annotation record is represented by a rectangle glyph that is colored according to the TE family assigned to the position. There are three SODA components in the visualization, stacked vertically in the following order:
\begin{itemize}
    \item TEs annotated on the forward strand of the chromosome
    \item Simple tandem repeat annotations (black glyph on gray genome background)
    \item TEs annotated on the reverse strand of the chromosome
\end{itemize}

Immediately following the SODA components are two additional visual components, not built with SODA: (i) a legend describing the colors used in the visualization, and (ii) a tabular description of each annotation rendered above (the table is a fixture of the Dfam website, found at \url{https://www.dfam.org/search/annotations}; it is not found in the example on sodaviz.org). Within the SODA components, hovering over glyphs triggers a highlight effect along with an informational tooltip popup. Additionally, this figure demonstrates the way that SODA objects can interact with the surrounding website based on SODA callback functions: when a glyph is clicked by the user, the table highlights and scrolls to the row corresponding to the clicked glyph.

\subsection{RepeatMasker visualization}
\begin{figure*}[t]
\begin{center}
\includegraphics[width=\textwidth,keepaspectratio=true]{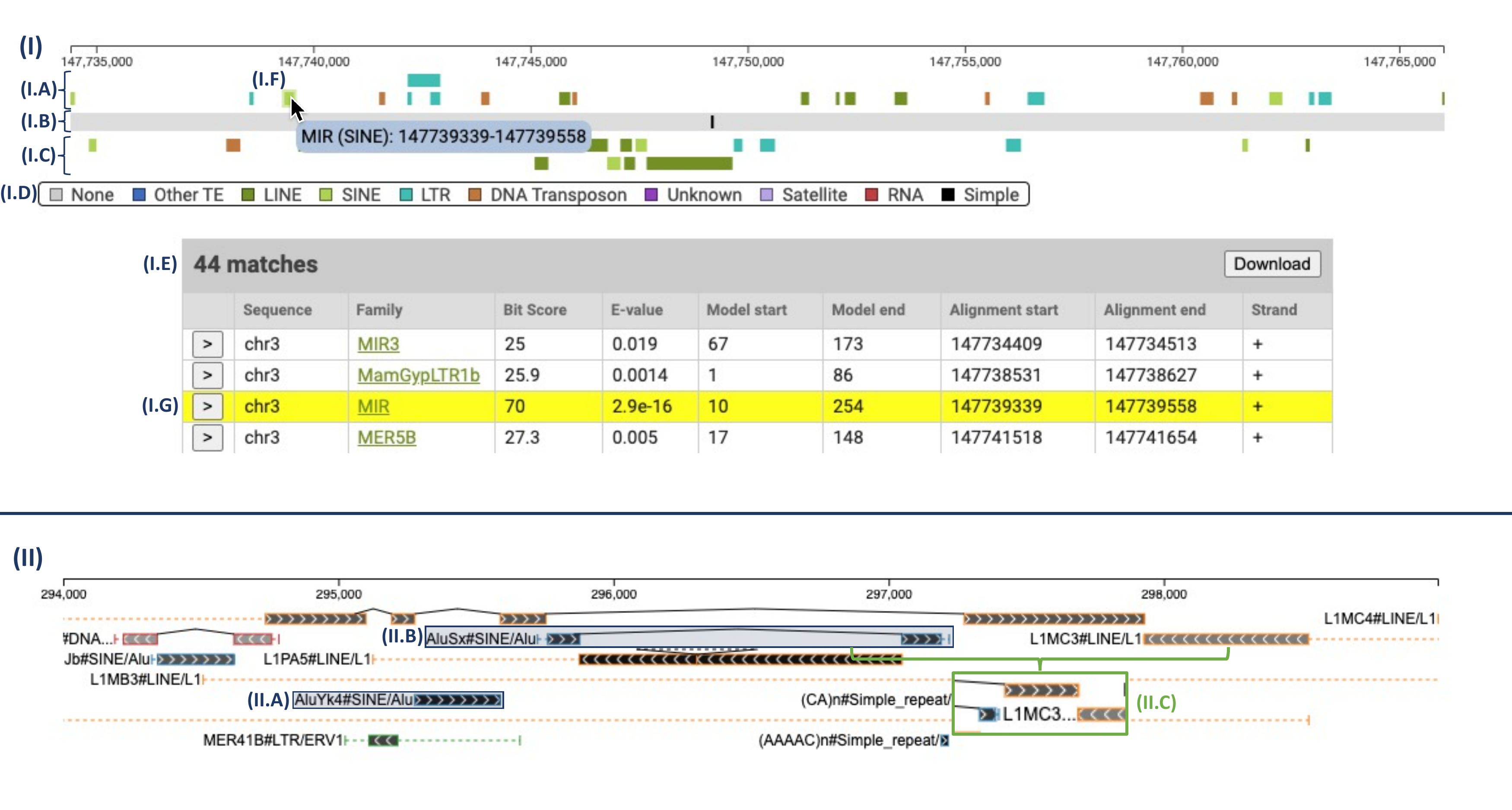}
\end{center}
\caption{(I) depicts the Dfam-SODA visualization. 
At (I.A), (I.B), and (I.C) are the forward strand TE annotations, simple repeat annotations, and reverse strand TE annotations, respectively.
At (I.D) is the repeat classification color map, and at (I.E) is the table describing the annotations visualized -- neither (I.D) nor (I.E) are created with SODA.
At (I.F), the green rectangle has been hovered and clicked, causing the descriptive tooltip to appear and the annotation's corresponding row at (I.G) to be highlighted.
(II) depicts the RepeatMasker-SODA visualization.
At (II.A) is a simple glyph representing a full-length TE annotation. At (II.B) is a compound glyph representing a joined group of TE fragment annotations.
At (II.C), an image is shown of the L1MC3 text label after the visualization is zoomed out -- when the label has less screen space, it renders a less detailed string.}
\label{fig:rmsk-dfam}
\end{figure*}

The UCSC genome browser~\cite{navarro2021ucsc} is a popular tool that houses visualization tracks for dozens of kinds of genomic annotation. The tracks are independently configured and many are submitted by external groups; they can differ greatly in visual complexity. The RepeatMasker Viz. track is a highly nuanced and information-dense track that represents the annotation of TEs and other repetitive DNA features as labeled by the tool RepeatMasker~\cite{smit2004repeat}. Using SODA, we developed a copy-cat implementation of the RepeatMasker Viz. track (Fig.~\ref{fig:rmsk-dfam}) to serve as a test-case with complex visual expression requirements.

The RepeatMasker Viz. track presents the same type of annotations as the Dfam visualization described in the previous section, but includes additional visual indicators to represent complex relationships that are not present in the Dfam visualization. As in the Dfam visualization, annotated TEs and simple repeats are represented with rectangle glyphs. The rectangles in the RepeatMasker plot, however, differ in three ways:

\begin{itemize}
    \item The outline, rather than the entire glyph, is colored according to the TE family color scheme.
    \item The interior of the rectangles are shaded in grayscale to indicate the inferred biological age (determined by the quality of the sequence alignment that defines the annotation) of the annotated feature. Younger features appear darker, while older features appear lighter.
    \item The interior of each rectangle is textured with a repeating chevron pattern to indicate the chromosome strand on which the feature is found.
\end{itemize}

TE annotations in a genome are often fragments of a full-length known TE family. These fragments can arise from partial replication (at time of insertion), partial deletion (some time after full-length insertion), or insertion of a newer element into the middle of a previously inserted TE (resulting in two fragments for the older element). Two indicators are added to the RepeatMasker visualization to represent these events:
\begin{itemize}
    \item When an annotation is identified as a fragment, dashed horizontal lines are rendered on either side to indicate portions of the consensus TE sequence that are missing from the fragment. 
    \item When discontiguous fragments are inferred to be the result of an older element having been split by a newer insertion, they are joined by two angled lines meeting at a point.
\end{itemize}

We refer to groups of joined fragments, dashed lines, and angled lines as a compound glyph.
A text label is placed immediately to the left of each compound glyph. As the user zooms in and out, the labels automatically adjust the level of displayed text detail, based on available space.

\subsection{PolyA visualization}
\begin{figure*}[t]
\begin{center}
\includegraphics[width=\textwidth,keepaspectratio=true]{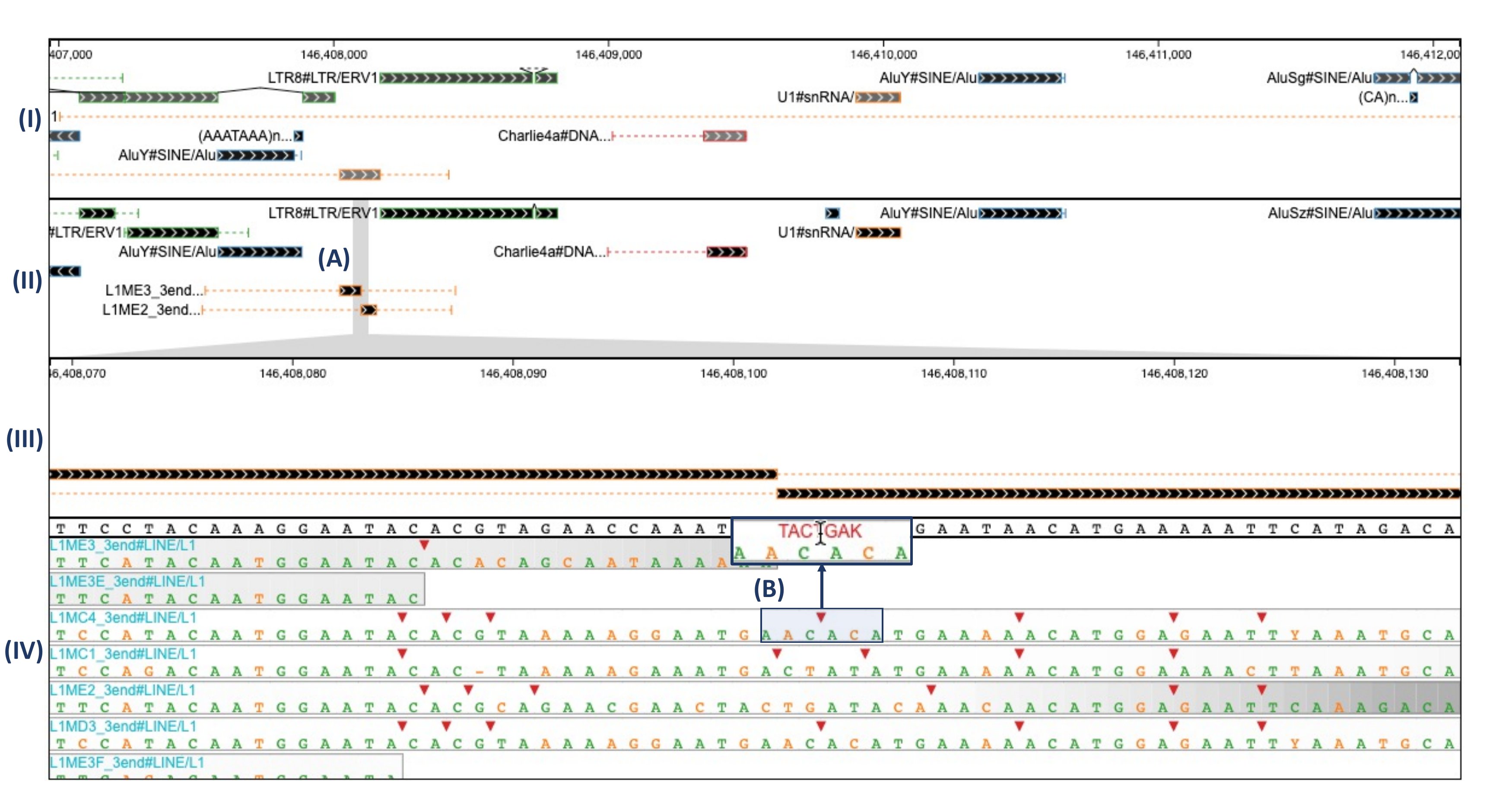}
\end{center}
\caption{In the PolyA-SODA visualization, pre-existing TE annotations are shown at (I) and PolyA-adjudicated annotations are shown at (II).
At (A), a brush selection has been made, and the selected interval is magnified and displayed at (III).
The alignments and confidence scores corresponding to the selected interval are shown at (IV).
(B) shows the results of hovering over one of the triangle glyphs that represent alignment insertions relative to the genome -- the inserted sequence replaces the triangle while hovered}
\label{fig:polya}
\end{figure*}

Our group has developed a tool, called PolyA~\cite{carey2021polya}, that adjudicates between competing alignment-based annotations by computing position-specific measures of confidence, identifying a trace with maximal confidence, and recursively splicing/stitching inserted elements. For debugging and exploration purposes, we found it helpful to develop a visualization application that provides insight into the differences between annotation results with PolyA and an alternative adjudication method (see Fig~\ref{fig:polya}).

Because PolyA's development was motivated by the goal of improving annotation of TEs, the visualization places PolyA information in the context of RepeatMasker's adjudication results, as produced by its internal tool, ProcessRepeats. Specifically, the PolyA-SODA visualization has four components:
\begin{itemize}
    \item A component displaying annotations queried from a mirror of the UCSC RepeatMasker database (these represent the results from the adjudication method found in RepeatMasker, called ProcessRepeats).
    \item A component displaying the PolyA adjudicated annotations for the same region. The user can make a brush selection on this component, resulting in changes to the following two components.
    \item A component that displays the exact (magnified) contents of the brushed interval in the above component.
    \item A component that displays a heatmap of the confidence scores for competing alignments in the same region. The sequence alignments from which the confidence scores were calculated are overlaid on top of the heatmap for each alignment.
\end{itemize}

This tool enables visual inspection of complicated genomic regions, identifying differences between the two annotation adjudication methods, and providing quick visual access to the sequence alignments and corresponding confidence values computed within PolyA.

\subsection{VIBES visualization}
We are currently developing a tool for the identification of phage sequences integrated into bacterial genomes, called VIBES (Viral Integrations in Bacterial gEnomeS).
To support data exploration by VIBES users, we have developed a SODA-based application consisting of several interconnected components.

The VIBES pipeline accepts a batch of bacterial genomes and a batch of phage genomes and identifies regions in the bacterial genomes that appear to be the result of phage integration. One output from VIBES is a representation, for each bacterial genome, of the locations of inferred (possibly partial) phage integrations -- there can be several per bacterial genome. Another VIBES output is, for each phage genome, a nucleotide-precision representation of the frequency with which each phage nucleotide is part of an observed integration across the batch of genomes. VIBES also identifies all Swiss-Prot and Pfam annotations for each phage genome.

\begin{figure*}[t]
\begin{center}
\includegraphics[width=\textwidth,keepaspectratio=true]{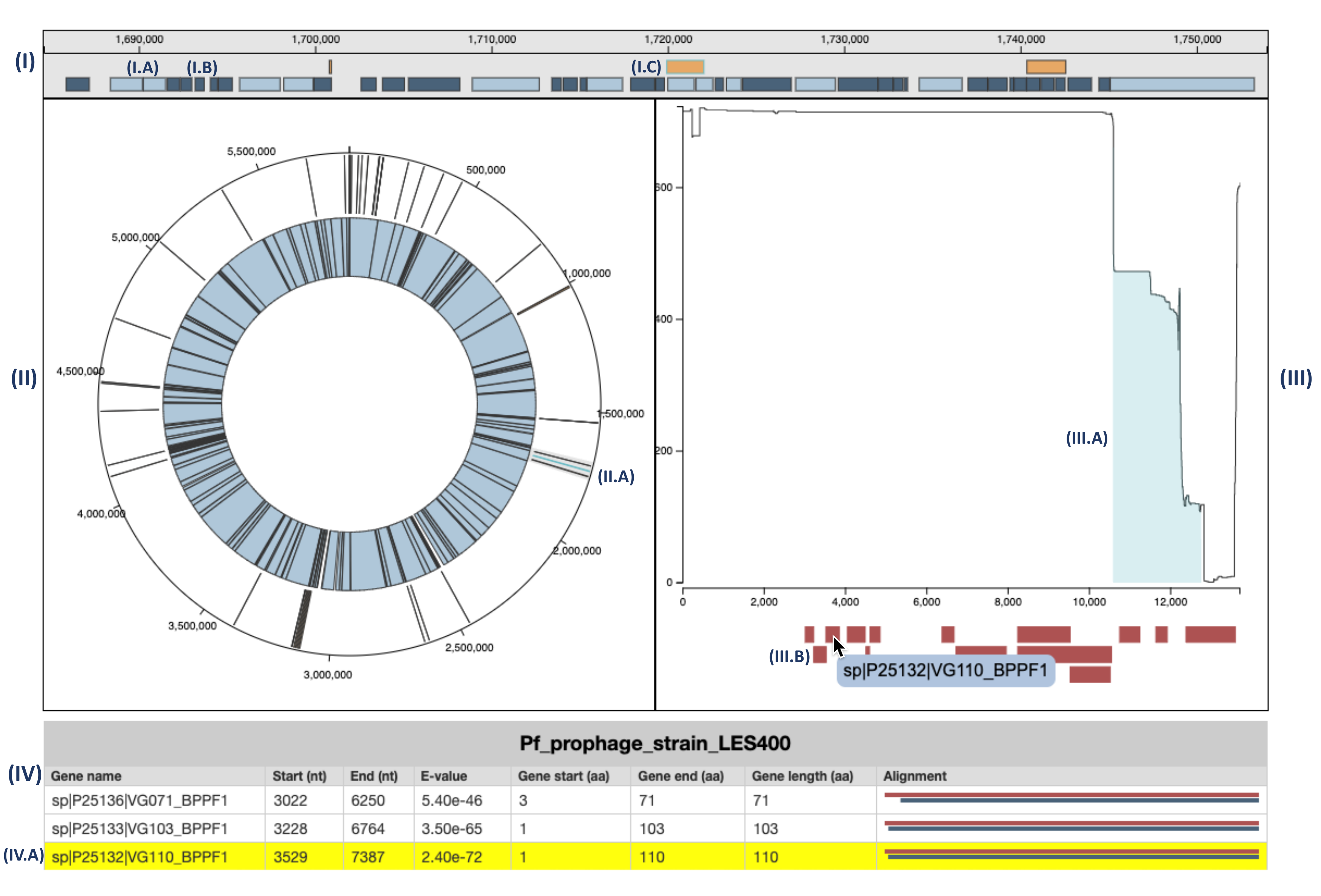}
\end{center}
\caption{In the VIBES-SODA visualization, the linear bacteria chart is displayed at (I).
At (I.A) is a handful of light blue bacterial gene annotations -- each of these rectangles represents an individual gene.
At (I.B) is a handful of dark blue bacterial gene annotations -- each of these rectangles represents a group of genes aggregated to simplify display.
At (I.C) is an orange phage integration annotation -- it is highlighted in cyan to indicate that it is the current \textit{selected} annotation.
At (II) is the circular bacteria chart.
At (II.A) we see the same selected phage annotation and the gray shading that indicates the region that is currently rendered in the linear chart.
At (III) is the occurrences plot, which displays position specific integration counts across all bacterial genomes for the selected phage integration annotation.
At (III.A) is the interval on the plot that indicates the portion of the phage genome that was identified in the selected phage annotation.
At (III.B) is the collection of viral gene annotations for the selected phage.
At (IV) is the table that describes the viral gene annotations shown in (III.B).
One of the viral gene annotations has been hovered and clicked, causing the tooltip at (III.B) and the row at (IV.A) to be highlighted.}
\label{fig:vibes}
\end{figure*}

At the top of the VIBES visualization (Fig.~\ref{fig:vibes}) is a chart that represents a linearized version of the target bacterial genome. The top row of the linear chart displays annotations of phage integrations to the bacterial genome as identified by VIBES, and the bottom row displays gene annotations. At the lowest zoom level, the gene annotations are condensed into groups based on proximity in the chromosome. As the zoom level increases, the gene annotations are continuously re-rendered as more horizontal screen real estate enables finer-grained separation and display.

Below the linear chart on the left is a circular representation of the bacterial genome. Here, phage integrations are shown on the outer ring and a condensed representation of gene annotations are on the inner ring. As the linear chart is zoomed and panned, a brush is drawn on the circular chart that highlights the region of the bacterial genome that is currently shown in the linear view.

On both the linear and circular charts, a single phage integration is highlighted blue, indicating that it is currently “selected”. The selected annotation determines the contents of the occurrences plot, which is directly to the right of the circular chart. The top of the occurrences plot displays the position-specific integration frequencies of the phage responsible for the current selected annotation. In the occurrences plot, the region of the phage genome that was identified in the current bacterial genome is highlighted in blue (a partial integration will have blue highlight over just the corresponding fragment of the full length phage). The bottom of the occurrences plot indicates positions of genes and protein domains along the entirety of the phage genome. The selected annotation can be switched by clicking on phage annotations on either the linear or circular chart, or by pressing the arrow buttons that appear when hovering over the circular chart.

At the bottom of the visualization is a simple table that describes the gene annotations across the selected phage genome. Clicking on an annotation in the phage occurrences plot will cause the corresponding entry to be highlighted in the table.

\section{DISCUSSION}

Here, we have introduced SODA, an open-source TypeScript/JavaScript library that enables the development of web-based biological sequence annotation visualizations. SODA is primarily intended for creating visualization systems, but it also has utility in the generation of one-off figures.
In particular, we imagine SODA as a replacement not for existing domain-specific tools, but for lower-level, general purpose web visualization technologies.
% SODA’s feature set is broadly capable, but is currently perhaps best utilized to create a baseline to be augmented with light use of general purpose libraries.
SODA’s feature set is broadly capable, but, in our experience, it is often useful to augment a SODA-based visualization by integration with components built with other web technologies.
We have designed SODA to be easy to use, broadly expressive, and extensible; we would be grateful for community engagement, in the form of identifying feature gaps and contributing to SODA's source code to extend expression options.    

\section{ACKNOWLEDGEMENTS}

We are grateful to Jeb Rosen and Robert Hubley for the extensive and valuable feedback they provided during development of the library, and Daniel Olson and Thomas Colligan for helpful comments on drafts of the manuscript. We thank Audrey Shingleton and Conner Copeland for their aid in development of example applications of SODA (PolyA and VIBES, respectively). This work was supported by NIH	grants R01-GM132600 and U24-HG010136, and institutional support from the University of Montana.

\subsubsection{Conflict of interest statement.} None declared.

%Bibliography
\bibliographystyle{unsrt}  
\bibliography{references}

\begin{thebibliography}{10}

\bibitem{buels2016jbrowse}
Robert Buels, Eric Yao, Colin~M Diesh, Richard~D Hayes, Monica Munoz-Torres,
  Gregg Helt, David~M Goodstein, Christine~G Elsik, Suzanna~E Lewis, Lincoln
  Stein, et~al.
\newblock J{B}rowse: a dynamic web platform for genome visualization and
  analysis.
\newblock {\em Genome biology}, 17(1):1--12, 2016.

\bibitem{howe2021ensembl}
Kevin~L Howe, Premanand Achuthan, James Allen, Jamie Allen, Jorge
  Alvarez-Jarreta, M~Ridwan Amode, Irina~M Armean, Andrey~G Azov, Ruth Bennett,
  Jyothish Bhai, et~al.
\newblock Ensembl 2021.
\newblock {\em Nucleic acids research}, 49(D1):D884--D891, 2021.

\bibitem{navarro2021ucsc}
Jairo Navarro~Gonzalez, Ann~S Zweig, Matthew~L Speir, Daniel Schmelter, Kate~R
  Rosenbloom, Brian~J Raney, Conner~C Powell, Luis~R Nassar, Nathan~D Maulding,
  Christopher~M Lee, et~al.
\newblock The {U}{C}{S}{C} genome browser database: 2021 update.
\newblock {\em Nucleic acids research}, 49(D1):D1046--D1057, 2021.

\bibitem{lyi2021gosling}
Sehi L'Yi, Qianwen Wang, Fritz Lekschas, and Nils Gehlenborg.
\newblock Gosling: A grammar-based toolkit for scalable and interactive
  genomics data visualization.
\newblock {\em IEEE Transactions on Visualization and Computer Graphics},
  28(1):140--150, 2021.

\bibitem{watkins2017protvista}
Xavier Watkins, Leyla~J Garcia, Sangya Pundir, Maria~J Martin, and UniProt
  Consortium.
\newblock {ProtVista: visualization of protein sequence annotations}.
\newblock {\em Bioinformatics}, 33(13):2040--2041, 03 2017.

\bibitem{bostock2011d3}
Michael Bostock, Vadim Ogievetsky, and Jeffrey Heer.
\newblock D3 data-driven documents.
\newblock {\em IEEE Transactions on Visualization and Computer Graphics},
  17(12):2301–2309, dec 2011.

\bibitem{pixijs}
Pixijs.
\newblock \url{https://pixijs.com/}.

\bibitem{storer2021dfam}
Jessica Storer, Robert Hubley, Jeb Rosen, Travis~J Wheeler, and Arian~F Smit.
\newblock The {D}fam community resource of transposable element families,
  sequence models, and genome annotations.
\newblock {\em Mobile DNA}, 12(1):1--14, 2021.

\bibitem{smit2004repeat}
Arian~FA Smit.
\newblock {R}epeat-{M}asker {O}pen-3.0.
\newblock {\em http://www. repeatmasker. org}, 2004.

\bibitem{carey2021polya}
Kaitlin~M Carey, Robert Hubley, George~T Lesica, Daniel Olson, Jack~W Roddy,
  Jeb Rosen, Audrey Shingleton, Arian~F Smit, and Travis~J Wheeler.
\newblock Poly{A}: a tool for adjudicating competing annotations of biological
  sequences.
\newblock {\em bioRxiv}, 2021.

\end{thebibliography}

\end{document}